\documentclass[aps,prb,reprint, superscriptaddress]{revtex4-2}
\usepackage[english]{babel}
\usepackage[utf8]{inputenc}

\usepackage{amsmath}
\usepackage{amsthm}
\usepackage{mathtools}
\usepackage{graphicx}
\usepackage{amsfonts}
\usepackage{amssymb}
\usepackage{bbm}
\usepackage{hyperref}
\usepackage{listings}
\usepackage{xcolor}
\usepackage{setspace}
\usepackage{braket}
\newcommand{\transp}{\mathsf{T}}
\renewcommand{\vec}[1]{\boldsymbol{\mathbf{#1}}}
\DeclareMathOperator*{\argmin}{arg\,min}
\DeclareMathOperator*{\argmax}{arg\,max}
\newcommand{\norm}[1]{\left\lVert#1\right\rVert}
\newcommand{\transpc}{\transp}
\begin{document}
	
	\title{Connecting Tikhonov regularization to the maximum entropy method for the analytic continuation of quantum Monte Carlo data}
	\author{Khaldoon Ghanem}
	\affiliation{Quantinuum, Leopoldstrasse 180, 80804 Munich, Germany}
	
	\author{Erik Koch} 
	\affiliation{J\"ulich Supercomputer Centre, Forschungszentrum J\"ulich, 52425 J\"ulich, Germany}
	\affiliation{JARA High-Performance Computing, 52425 J\"ulich, Germany}
	
	\date{\today}
	
	\begin{abstract}
		Analytic continuation is an essential step in extracting information about the dynamical properties of physical systems from quantum Monte Carlo (QMC) simulations.
		Different methods for analytic continuation have been proposed and are still being developed.
		This paper explores a regularization method based on the repeated application of Tikhonov regularization under the discrepancy principle. 
		The method can be readily implemented in any linear algebra package and gives results  surprisingly close to the maximum entropy method (MaxEnt).
		We analyze the method in detail and demonstrate its connection to MaxEnt.
		In addition, we provide a straightforward method for estimating the noise level of QMC data, which is helpful for practical applications of the discrepancy principle when the noise level is not known reliably.
	\end{abstract}
	
	\maketitle
	
	\section{Analytic continuation: an ill-posed problem}
	From a mathematical perspective, the analytic continuation problem corresponds to solving a Fredholm integral equation of the first kind
	\begin{equation}\label{eq:fredholm}
		g(y) = \int dx K(y, x) f(x)\;,
	\end{equation}
	where $f(x)$ is the unknown spectrum, a non-negative integrable function. $K(y, x)$ is the kernel of the integral equation and is known analytically, while $g(y)$ is noisy data, typically obtained from QMC simulation at a finite number of points $y_j$.
	
	To solve the analytic continuation numerically, the integral  is discretized using a grid of $n$ points $x_i$,  giving a linear system of equations
	\begin{equation}\label{eq:system}
		\vec{g} = \vec{K} \vec{f}\;,
	\end{equation}
	where the elements of the matrix $\vec{K}$ are the kernel values $K(y_j, x_i)$, $\vec{g}$ contains $m$ measured data values $g(y_j)$ and ${f}_i$ is the spectrum integral over the i-th grid interval.
	The most naive and straightforward way of  solving Eq.~\eqref{eq:system}  is, as with any other linear system of equations, using the weighted least squares method
	\begin{equation}\label{eq:ls}
		\vec{f_\text{LS}}=
		\underset{\vec{f}}{\argmin}\ \chi^2(\vec{f}) \;,
	\end{equation}
	which finds the spectrum minimizing the fit to the data
	\begin{equation}
		\chi^2(\vec{f})\coloneqq  \left(\vec{g}  -\vec{K} \ \vec{f} \right)^\transpc \vec{C}^{-1} \left(\vec{g} - \vec{K} \ \vec{f} \right)\;.
	\end{equation}
	The fit is weighted by the inverse of $\vec{C}$, the covariance matrix of the noise on the data. By factorizing the covariance matrix into $\vec{C}^{-1} = \vec{T}^\transpc \vec{T}$, 
	one can always replace the kernel matrix and data vector by the weighted ones $\vec{T} \vec{K}$ and $\vec{T} \vec{g}$, respectively.
	Then the covariance matrix of the weighted data becomes the identity matrix, and one can use the ordinary least squares method instead.
	In the following, we will always assume that such transformation has been applied to the kernel and the data despite using the same notation $\vec{K}$ and $\vec{g}$ to denote the weighted ones.
	
\begin{figure}[t]
		\center
		\includegraphics[width=\columnwidth]{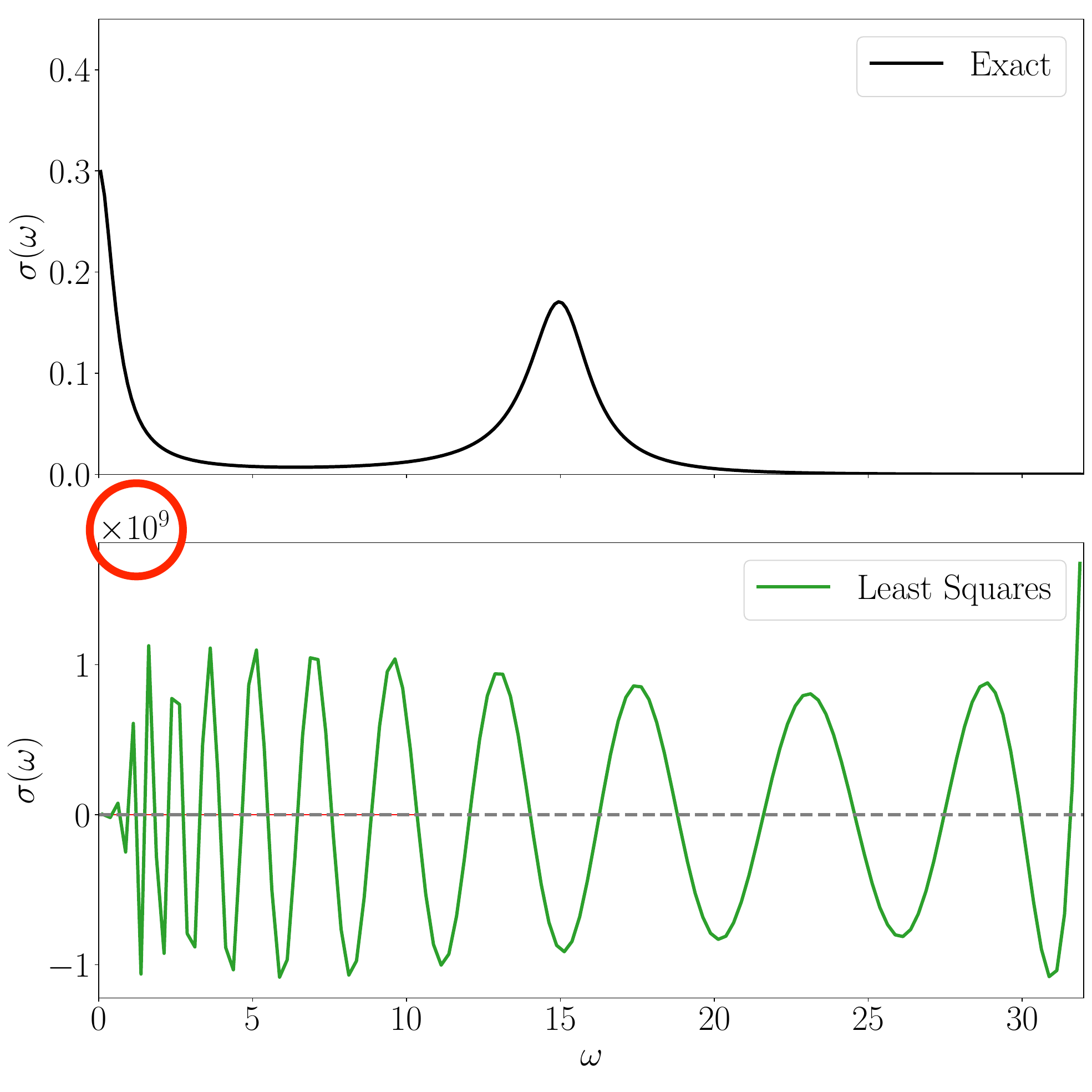}
		\caption{\label{fig:ls} 
			Least squares solution (bottom panel) for the analytic continuation of optical conductivity $\sigma(\omega)$ using noisy data of its correlation function.
			The exact correlation function is computed analytically from the exact optical conductivity (top panel) on the first $m=60$ bosonic Matsubara frequencies with inverse temperature $\beta=15$.
			The input data includes relative Gaussian noise with standard deviation $10^{-2}$.
			This test case is an adaptation of the ones proposed by Ref.~\cite{Gunnarsson10b} and studied further in Refs.~\cite{Ghanem17, Ghanem20a, Ghanem20b}. 
			In the notation of the latter reference, the optical conductivity used here differs in the values of the following parameters: $\Gamma_e=20, \epsilon_1 = 15$.
			We denote this data set as \emph{test case 1}.
		}
	\end{figure}

\begin{figure}[t]
		\center
		\includegraphics[width=\columnwidth]{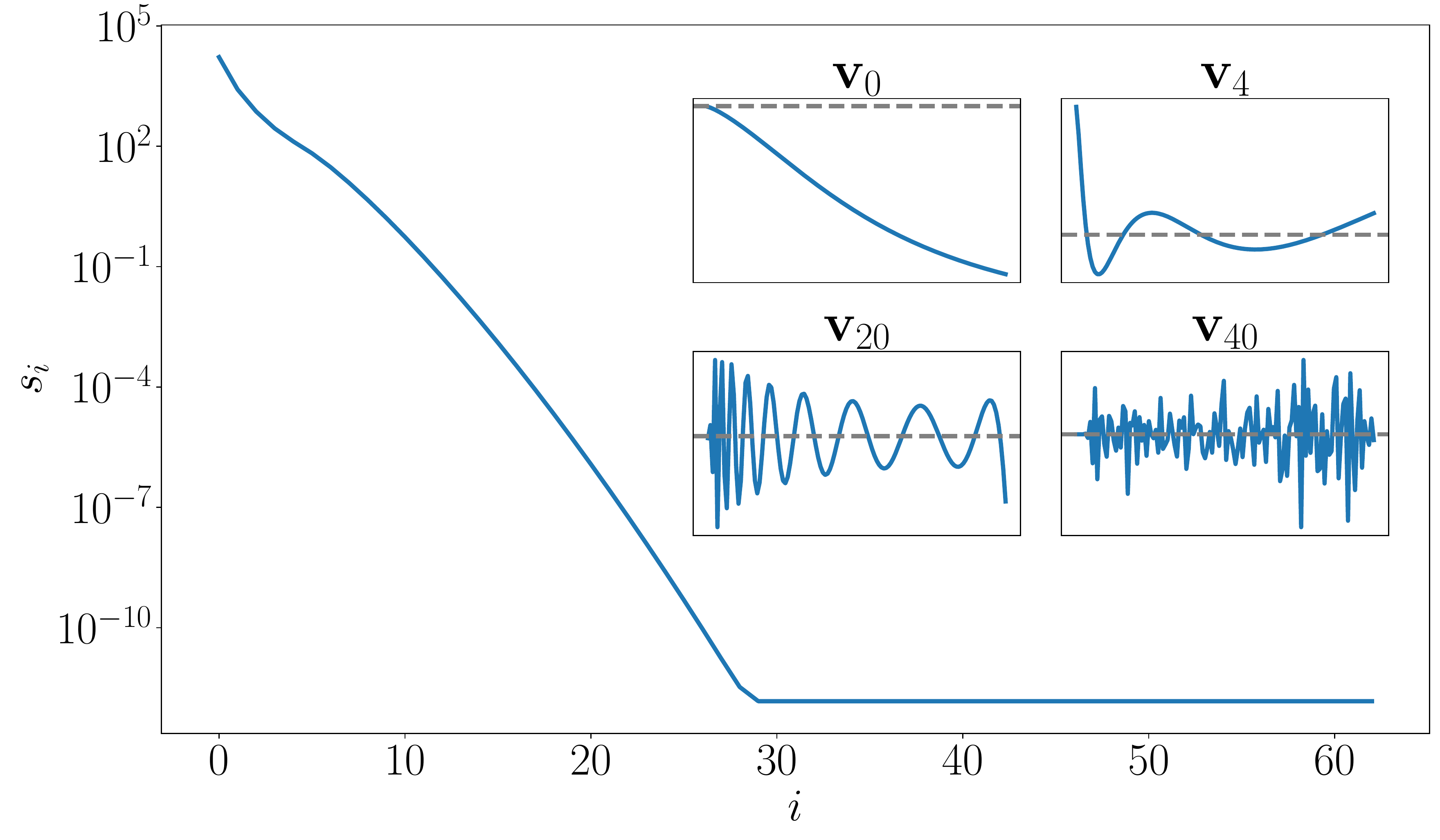}
		\caption{\label{fig:svd} 
			Singular values of the (weighted) kernel of test case 1.
			The singular values decay exponentially until leveling off at a value determined by the machine epsilon.
			In the inset, we show some of the spectral modes. The leading spectral modes are smooth and slowly varying functions.
			As the mode index increases, the number of nodes increases, and the modes become more oscillatory. 
			Once the singular values reach numerical accuracy, the corresponding modes become numerically degenerate so that the SVD routine returns arbitrary linear combinations of the exact modes.
		}
	\end{figure}

	Using the least squares solution for solving the analytic continuation problem gives generally bad results plagued by noise, as exemplified in Fig.~\ref{fig:ls}. 
	The reason is that the matrices  in analytic continuation problems are highly ill-conditioned such that the inevitable small noise on the data  leads to disastrous noise on the least-squares solution~\cite{Hansen92, Hansen10}.
	This can be seen more explicitly using the singular value decomposition (SVD) of the kernel matrix 
	\begin{equation}
		\vec{K} = \vec{U} \vec{S} \vec{V}^\transp\;,
	\end{equation}
	where $\vec{S}$ is a diagonal matrix of size $m \times n$, and $\vec{U}$ and $\vec{V}$ are unitary matrices of sizes $m \times m$ and $n \times n$, respectively. 
	The columns of the matrix $\vec{U}$ form an orthonormal basis of the data space and are called the \emph{data modes}, while the columns of the matrix $\vec{V}$,  which span the space of spectra, are called the \emph{spectral modes}.
	The diagonal elements of $\vec{S}$ are the \emph{singular values}, and they are sorted in descending order.
	Using the SVD, the least squares solution can be written as
	\begin{equation}\label{eq:ls_svd}
		\vec{f}_\text{LS} = \sum_{i}^{\min (m,n)} \frac{\vec{u}_i^\transpc \vec{g}}{s_i}\ \vec{v}_i\;.
	\end{equation}
	For matrices arising from analytic continuation problems, the singular values decay exponentially to zero (see Fig.~\ref{fig:svd}).
	Dividing by these vanishing singular values hugely amplifies any small noise present in the data. This is the main problem with the least-squares solution. 
	
	The other source of ill-posedness is the incompleteness of the data, i.e., we only know the data at a finite number of points $m<n$, where $n$ is typically chosen large enough to resolve the desired features of the spectrum. 
	Therefore, even for numerically exact data, if no regularization/additional information is provided, one can only ever hope to recover at most the first $m$ modes of the spectrum.
	
	\section{Noise estimation}
	The SVD of the kernel matrix allows an accurate estimation of the overall scale of noise on QMC data.
	This can be valuable in practical situations where such an estimate is unavailable, or as an important cross-check of the validity of the noise level estimate.
	
	As a start, let us assume, as usual, that an estimate of the covariance matrix $\vec{C}$ already exists and that the data and kernel have been weighted by $\vec{T}$, the square root of its inverse. 
	Consequently, the noise on the different components of the weighted data vector $\vec{g}$ is uncorrelated and has a unit variance.
	Since the matrix $\vec{U}$ is unitary, the noise $\epsilon_i$ present in the expansion coefficients of the data $	\vec{u}^\transpc_i \vec{g} $
	is also uncorrelated and has a unit variance.
	These noisy data coefficients are then related to the exact spectrum via the relation
	\begin{equation}
		\vec{u}^\transpc_i \vec{g} = s_i \ \vec{v}^\transp_i \vec{f}_\text{exact} + \epsilon_i\;.
	\end{equation}
	Given that the exact spectrum has a finite norm and that the singular values in analytic continuation decay exponentially, there is some index $k$, after which the exact data coefficients become negligible compared to the noise.
	For these indices, the measured data coefficients are practically plain noise
	\begin{equation}
		\vec{u_i}^\transpc\vec{g}\approx \epsilon_i \qquad : k<i\leq m \;,
	\end{equation}
	and can be used to estimate the variance of the noise $\epsilon_i$ as 
	\begin{equation}\label{eq:variance}
		\sigma^2(\epsilon) \approx {{\frac {1}{m-k}}\sum _{i=k+1}^{m}(\vec{u_i}^\transpc\vec{g}})^{2}\,,
	\end{equation}
	where the formula for estimating  population variance with a known mean of  value zero has been employed.
	In practice, the cutoff $k$ can be safely chosen as the numerical rank of $\vec{K}$, i.e., the index at which the singular values hit numerical accuracy.
	
\begin{figure}[t]
		\center
		\includegraphics[width=\columnwidth]{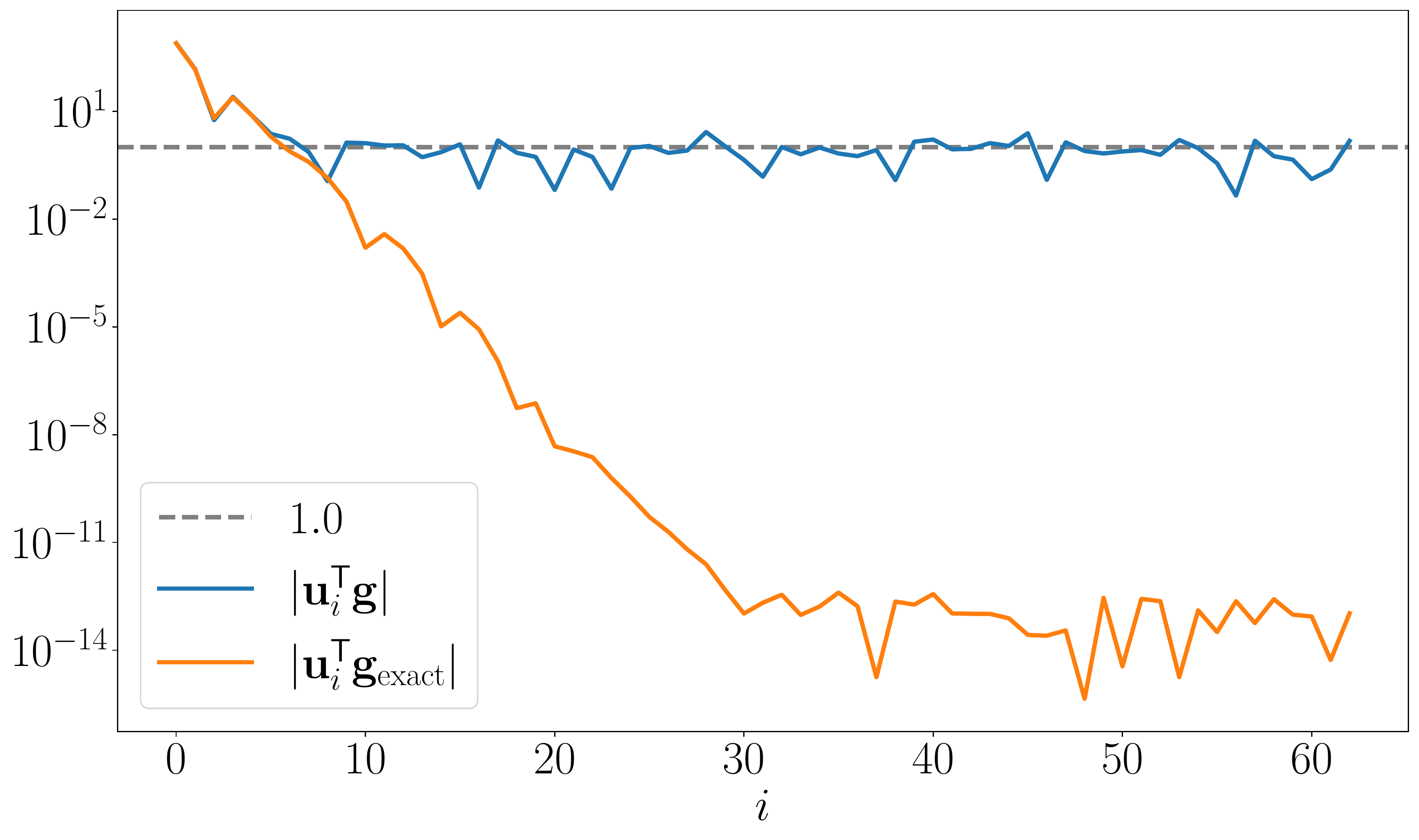}
		\caption{\label{fig:noise} 
			Absolute values of the exact and noisy data coefficients of test case 1. 
			While the exact coefficients decay to the machine epsilon, the noisy ones decay until they hit the noise level and then fluctuate around it.
			Here the noise level equals one because the data is weighted by the proper covariance matrix.
			Notice that large noisy coefficients are close to their exact values and that the deviation becomes significant only when their values drop to near the noise level.
		}
	\end{figure}

	When the covariance matrix $\vec{C}$ is properly scaled, we expect this value to be close to one.
	This is illustrated in Fig.~\ref{fig:noise} for test case 1, where the data coefficients decay exponentially till they reach the noise level $\sigma(\epsilon) = 1$ and fluctuate around it.
	However, when a covariance matrix with the wrong scaling is used, the aforementioned plateau of data coefficients will be scaled accordingly, and $\sigma(\epsilon)$ will deviate from the expected value of one. 
	Values much larger than one indicate that the noise level has been underestimated, while values much lower than one indicate an overestimation of the noise level.
	
	An important practical use case of the above formula is estimating the noise level of uncorrelated \emph{relative} Gaussian noise.
	In this case, as an initial Ansatz, one can use a diagonal covariance matrix whose diagonal elements are the squares of the data values.
	Eq.~\eqref{eq:variance} then provides an estimate of $\sigma^2$, which can be multiplied by the ansatz to obtain a properly-scaled covariance matrix.
	
	\section{Tikhonov regularization}
	The expansion of the least squares solution using SVD modes [cf. Eq.~\eqref{eq:ls_svd}] already suggests a direct remedy to the ill-posedness; namely, truncating the later modes, which are dominated by noise, while keeping the leading ones that are more stable.
	This is known as the Truncated SVD solution.
	Tikhonov regularization~\cite{Phillips62, Tikhonov77} is a more refined method, where the noisy modes are turned off continuously, with each term in the least squares solution multiplied by a filtering  function $\phi(s; \alpha) \coloneqq {s^2}/\left({s^2+\alpha}\right)$ that depends on its singular value $s$ and an adjustable parameter $\alpha$:
	\begin{equation}\label{eq:btikh_svd}
		\vec{f}_\text{Tikhonov}(\alpha) =\sum_{i}^{\min (m,n)} \phi(s_i; \alpha) \ \frac{\vec{u}_i^\transpc \vec{g}}{s_i}\ \vec{v}_i\;.
	\end{equation}
	Terms corresponding to very small singular values $s_i^2 \ll \alpha$ are practically removed, while ones corresponding to large singular values $s_i^2 \gg \alpha$ are hardly modified~\footnote{In the inverse-problem literature, it is common for Tikhonov regularization parameter $\alpha$ to appear squared. We choose to deviate from that convention in order to make the correspondence with the regularization parameter of MaxEnt more seamless.}.
	
	It can be shown that the above Tikhonov solution is the least squares solution of an alternative problem with extended data and an extended kernel
	\begin{equation}\label{eq:btikh_ls}
		\vec{f}_\text{Tikhonov}(\alpha) = \underset{\vec{f}}{\argmin} \norm{
			\begin{pmatrix} \vec {K} \\ \sqrt{\alpha}\ \vec{I}
			\end{pmatrix}  \vec{f} - \begin{pmatrix} \vec {g} \\ \vec{0}
		\end{pmatrix}  }^2\;,
	\end{equation}  
	where $\vec{I}$ is the unit matrix in the $n$-dimensional space of spectra.
	This formulation has  a computational advantage for large-scale problems because it allows getting the Tikhonov solution using any linear solver without explicit computation of the singular value decomposition.
	Moreover, this least squares problem can be written as the following minimization problem
	\begin{equation}\label{eq:btikh}
		\vec{f}_\text{Tikhonov}(\alpha)= \underset{\vec{f}}{\argmin}\ \chi^2(\vec{f}) + \alpha \norm{\vec{f}}^2\;, 
	\end{equation}
	that aims to balance the fit to the data with the $L_2$-norm of the spectrum vector.
	The balance is controlled by the regularization parameter $\alpha$. When $\alpha$ is very small, we approach the least squares solution, which fits the data very well but has a very large $L_2$-norm. 
	As $\alpha$ increases, more modes get filtered, and the norm gets smaller while the fit gets worse.
	The smoothness typically associated with Tikhonov solutions comes from the fact that the leading modes are smoother than later ones for analytic continuation kernels (see, for example, the insets of Fig.~\ref{fig:svd}).
	
	While the aforementioned form of Tikhonov regularization is the most basic and widely used one in the inverse problem literature \footnote{The most general form of Tikhonov is obtained by replacing the  $L_2$-norm with a bilinear function $\norm{\mathbf{f}-\mathbf{f}_0}^2_\mathbf{M}$, where $\mathbf{M}$ is some positive-definite matrix and $\mathbf{f}_0$ is an arbitrary vector that acts as a default model.},   it has two drawbacks for analytic continuation problems.
	The first is that the discretized $L_2$-norm is grid-dependent because the spectral values $f_i \coloneqq w_i f(x_i)$ include the full weight of the grid interval at point $x_i$.
	Using a grid with $n$ points and a grid density $\rho(x)$, these weights are defined as $w_i \coloneqq 1/\left[N \rho(x_i) \right]$ and the $L_2$- norm of the spectrum reads
	\begin{equation}
		\norm{\vec{f}}^2 = \sum_i f_i^2 = \sum_i \left[w_i f(x_i)\right]^2 \approx \frac{1}{N} \int dx\  \frac{f^2(x)}{\rho(x)}\;.
	\end{equation}
	This shows that the basic form of Tikhonov has an implicit dependence on the grid density \footnote{One can obtain trivial grid independence by including one square root of the grid weights in the spectrum vector and the other square root in the kernel matrix.
		In this case, the discretized  $L_2$-norm of $\vec{f}$ corresponds to the continuous $l_2$-norm of $f(x)$.
		However, using this form implies a specific choice of the measure on $x$ that is equivalent to fixing the grid density $\rho(x)$ to be uniform.}.
	We suggest replacing this implicit dependence with an explicit one on a default model $d(x)$.
	Let $d_i \coloneqq w_i d(x_i)$ be the integral of the default model over the $i$-th grid interval, then we replace the usual $L_2$-norm $\sum_i f_i^2$ with the weighted $L_2$-norm $\sum_i f_i^2/d_i$.
	It can be easily verified that the weighted norm is indeed grid-independent. 
	
	The second drawback is that the solution approaches zero in the limit of large regularization parameter $\alpha$.
	In analytic continuation, however, we know that the spectrum must have a finite $L_1$-norm, so it would be desirable if the solution would approach some properly normalized spectrum in the limit of large $\alpha$.
	We choose to center our regularization term at the default model $\vec{d}$  instead of zero.
	
	In summary, we propose using the following  form of Tikhonov regularization in analytic continuation problems
	\begin{equation}\label{eq:tikh}
		\vec{f}_\text{Tikhonov}(\alpha, \vec{d}) = \underset{\vec{f}}{\argmax} -\frac{1}{2}\chi^2\left(\vec{f}\right) + \alpha T\left(\vec{f}|\vec{d}\right)\;,
	\end{equation}
	where the Tikhonov penalty term is defined as
	\begin{equation}\label{eq:tikh_penalty}
		T(\vec{f}|\vec{d}) = -\frac{1}{2} \sum_i \frac{\left(f_i -d_i \right)^2}{d_i}\;.
	\end{equation}
	It is worth noting that, like the original form, this new formulation  can be solved as  an extended least squares problem 
	\begin{equation}\label{eq:tikh_ls}
		\vec{f}_\text{Tikhonov}(\alpha, \vec{d}) = \underset{\vec{f}}{\argmin} \norm{
			\begin{pmatrix} \vec {K} \\ \sqrt{\alpha\ \vec{D}^{-1}}
			\end{pmatrix}  \vec{f} - \begin{pmatrix} \vec {g} \\ \sqrt{\alpha \vec{D} }\vec{e}
		\end{pmatrix}  }^2\;,
	\end{equation}  
	with $\vec{D}=\text{diag}(\vec{d})$ and $\vec{e} \coloneqq \left(1, 1, \dots, 1\right)^\transp$. Its solution can  be similarly expressed in terms of the SVD of the rescaled kernel $ \vec{K} \sqrt{\vec{D}}$ as shown in appendix~\ref{app:tikh}.
	
	\section{Discrepancy principle}\label{sec:discrepancy}
	Choosing the value of the regularization parameter $\alpha$ is an essential ingredient of any regularization method. 
	Apart from the obvious criterion that $\alpha$ should be smaller for more accurate data, there is no unique procedure for actually determining its value.
	Any such procedure should strike a balance between fitting the noise and biasing the solution.
	A common method in the inverse problem literature is the discrepancy principle \cite{groetsch1984,Morozov1984}. 
	
	According to the discrepancy principle, a good spectrum would produce data such that the residual vector $\vec{r} \coloneqq \vec{g} - \vec{K} \vec{f}$ is dominated by noise. Therefore, we should choose $\alpha$ such that the norm of the residual $\norm{\vec{r}}^2 = \chi^2(\vec{f})$ equals the expected norm of the noise vector.
	Assuming, as usual, that data and kernel have been reweighed with the square root of the noise covariance, the expected norm-squared of the noise vector follows the well-known chi-squared distribution.
	The mean value of this distribution equals the number of data points $m$, and its variance equals $2m$.
	To avoid accidental over-fitting of noise, one may apply the discrepancy principle using a value (in terms of the standard deviation) somewhat larger than the mean.
	In this work, however, we always use the mean value.

	Interestingly, the Tikhonov solution using the discrepancy principle can be written  in a form independent of any regularization parameter $\alpha$ as a maximization of the Tikhonov penalty
	\begin{equation}\label{eq:tikh_disc}
		\vec{f}_\text{Tikhonov}(\vec{d}) =\underset{\vec{f} \in \mathcal{C}} {\argmax}\; T\left(\vec{f}|\vec{d}\right)\;,
	\end{equation}
	over the manifold $\mathcal{C}$ defined by the discrepancy principle
	\begin{equation}\label{eq:disc_set}
		\mathcal{C} \coloneqq \left\{\vec{f} \in \mathbb{R}^n : \chi^2(\vec{f}) = m \right\}\;.
	\end{equation}
	Starting from some spectrum on the manifold $\mathcal{C}$, the Tikhonov solution can then be found by following the gradient of $T\left(\vec{f}|\vec{d}\right)$, projected on $\mathcal{C}$:
	\begin{equation}
		\vec{a}^{\perp} = \left[ \vec{I}- \frac{ \vec{z}\; \vec{z}^\transpc }{\vec{z}^\transpc\vec{z}}\right] \vec{a} \;,
	\end{equation}
	where $\vec{a} \coloneqq \nabla T$ is the gradient of the Tikhonov penalty with
	\begin{equation}\label{eq:tikh_grad}
		a_i  = - \frac{f_i-d_i}{d_i}\;,
	\end{equation}
	and $\vec{z} \coloneqq -\frac{1}{2}\nabla \chi^2$ is the gradient of the fit function  i.e. the surface normal of $\mathcal{C}$ with
	\begin{equation}
		z_i = \vec{k}_i^\transpc \left[\vec{g} - \vec{K} {\vec{f}}\right] \;,
	\end{equation}
	where $\vec{k}_i$ is the $i$-th column of the Kernel matrix $\vec{K}$.
	At the optimal point, the projection vanishes, and the gradient of $T$ must be anti-parallel to the fit gradient
	\begin{equation}\label{eq:tikh_stationarity}
		\alpha \ \vec{a} = - \vec{z}\;,
	\end{equation}
	which is nothing but the stationarity condition for \eqref{eq:tikh}.
	The optimal regularization parameter $\alpha$ thus reemerges as the ratio of the two gradients at the optimal point.
	
	In practice, this constrained optimization problem is converted, using the method of Lagrange multiplier, into an unconstrained optimization of the objective function 
	\begin{equation}\label{eq:tikh_lagrange}
		\vec{f}_\text{Tikhonov}(\vec{d}) =\underset{\vec{f}, \beta} {\argmax}\; T\left(\vec{f}|\vec{d}\right) - \frac{\beta}{2}\left[ \chi^2(\vec{f}) - m\right]\;,
	\end{equation}
	where the Lagrange multiplier $\beta$ corresponds to the inverse of the regularization parameter $\alpha$.
	
	\section{Self-Consistent Tikhonov}

\begin{figure}[t]
		\center
		\includegraphics[width=\columnwidth]{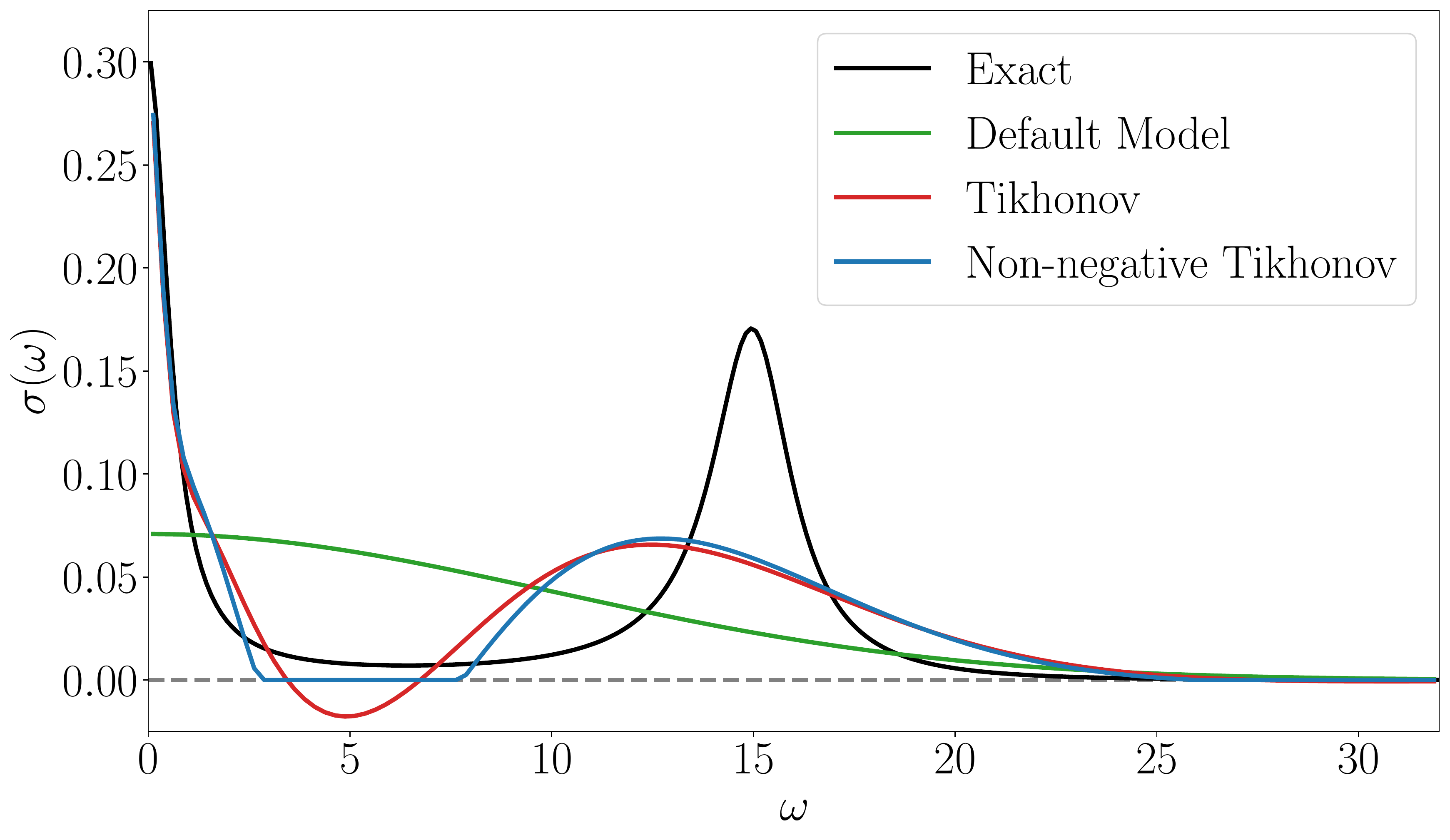}
		\caption{\label{fig:tikh1} 
			Tikhonov solutions for test case 1 using a Gaussian default model centered at $0$ with width $10$.
			The values used for the regularization parameter $\alpha$ are determined by the discrepancy principle.
		}
	\end{figure}

	Tikhonov regularization provides a simple and fast method to obtain a decent first impression of the analytic continuation solution.
	Its obvious disadvantage, however, is ignoring the non-negativity of the spectrum (see Fig.~\ref{fig:tikh1}).
	One can enforce the non-negativity by explicitly restricting the optimization problem to non-negative spectra.
	This can be done straightforwardly by using the non-negative least squares method~\cite{Lawson95} with the extended kernel and data of Eq.~\eqref{eq:tikh_ls}.
	Nevertheless, enforcing the non-negativity in this artificial way does not improve the results as desired.
	As shown in Fig.~\ref{fig:tikh1}, the non-negative Tikhonov solution looks like a clamped version of the original Tikhonov solution where the negative parts are set to zero, while the positive part stays roughly the same with minor adjustments to account for the truncated negative values.
	
	Instead of enforcing the non-negativity constraint directly, one can  reduce violations by increasing the regularization parameter $\alpha$, which encourages the solution to be close to the non-negative default model. Under the discrepancy principle, the regularization parameter is determined implicitly and only has a large value if the default model fits the data well.
	This transforms the problem of satisfying non-negativity into one of improving the fit of the default model.
	In the limit, when the default model itself satisfies the discrepancy principle, it is its own Tikhonov solution, and thus non-negativity is guaranteed.

	A simple way of improving the fit of a default model is by linearly mixing it with its Tikhonov solution under the discrepancy principle
	\begin{equation}
		\vec{d} \gets [1-\mu]\ \vec{d} + \mu\ \vec{f}_\text{Tikhonov}(\vec{d})\;.
	\end{equation}
	Assuming the fit of the starting default model is worse than $m$, the new default model is guaranteed to have a better fit due to the convexity of the fit function $\chi^2$.
	Additionally, if the starting default model is strictly positive, we can always choose the positive mixing parameter $\mu$ small enough such that the new default model is also positive.
	The values of the mixing parameter that guarantee the positivity of the new default model can be calculated explicitly from the values of the starting default model and its Tikhonov solution as
	\begin{equation}\label{eq:mu_plus}
		\mu < \min \left\lbrace \frac{d_i}{d_i - f_i} : f_i < d_i \right\rbrace\;.
	\end{equation}
	These observations suggest an iterative approach to obtain an improved non-negative Tikhonov solution.
	In this approach, we keep linearly mixing the default model with its Tikhonov solution to obtain a new, improved default model until the difference between the default model and its Tikhonov solution becomes negligible. We call
	this method \textbf{Self-Consistent Tikhonov (SCT)}.
	
	For the mixing parameter $\mu$, we use half the maximum allowed value  [cf. Eq.~\eqref{eq:mu_plus}].
	Using this value implies that the updated default model has at least half its original value at any point.
	This mixing strategy works well for most cases, but it can sometimes lead to slow convergence when the exact spectrum has values very close to zero (e.g., at the tail of a Gaussian peak).
	To accelerate the convergence of such cases, we put a lower limit on the mixing parameter $\mu$.
	This may lead to a violation of the positivity of the default model, which can be directly reinforced by truncating values lower than some positive threshold.
	It should be emphasized that these limits  are not strictly necessary, but help accelerate convergence in pathological cases.
	
\begin{figure}[t]
		\center
		\includegraphics[width=\columnwidth]{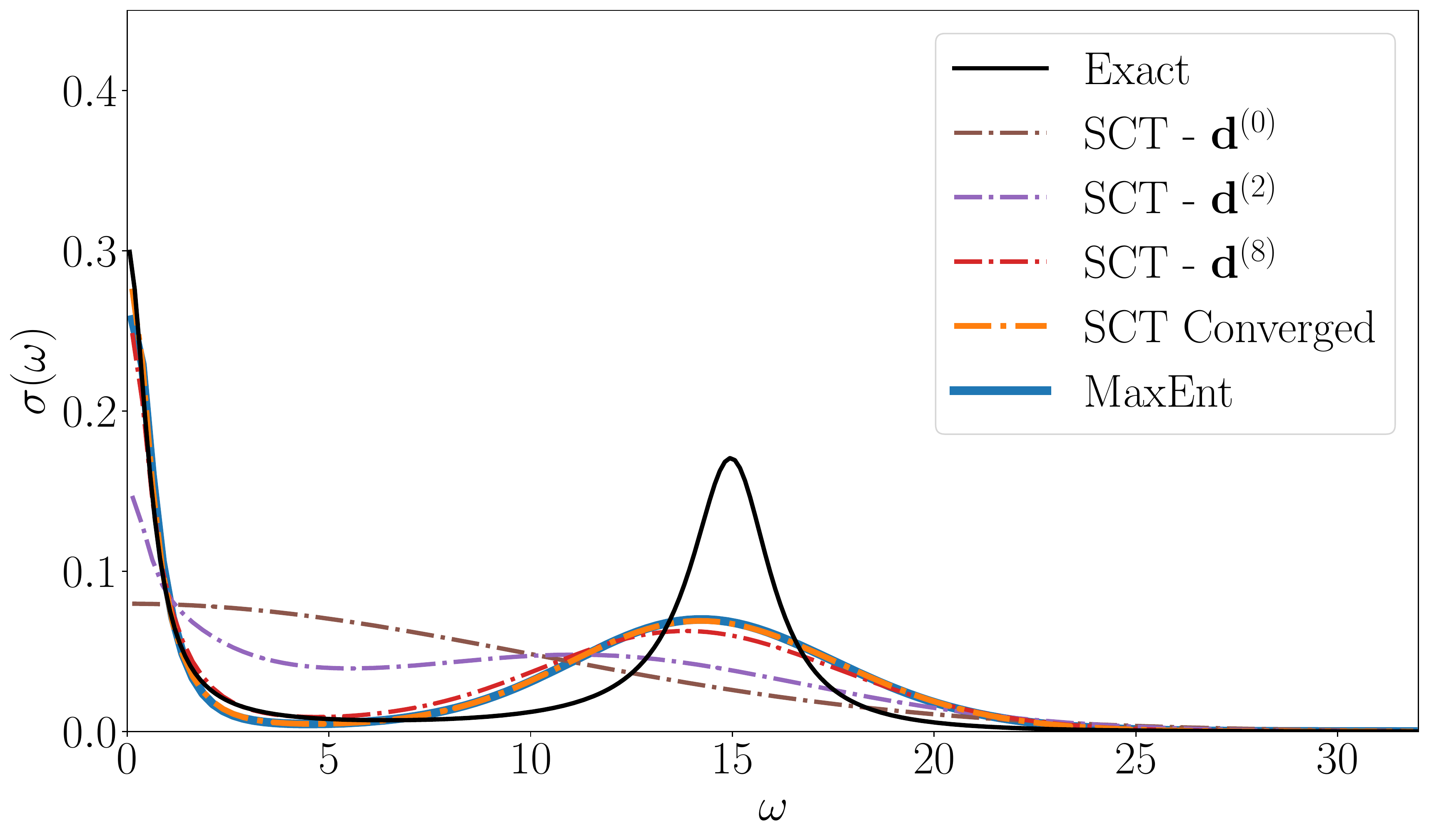}
		\caption{\label{fig:sct_dms} 
			Comparison of MaxEnt and default models produced by SCT at different iterations.
			The superscript of the default model represents its iteration number with 
			$\vec{d}^{(0)}$ being the starting default model.
			For MaxEnt, the starting default model $\vec{d}^{(0)}$ was used, and the regularization parameter was determined by the discrepancy principle.
		}
	\end{figure}

	In Fig.~\ref{fig:sct_dms}, we plot a set of default models produced by SCT for test case 1 at different iterations.
	The default model gradually transforms and fits the data till it converges, with the converged solution satisfying the discrepancy principle.
	This solution represents a significant improvement over the original Tikhonov solution and its non-negative counterpart (see Fig.~\ref{fig:tikh1}).
	Besides providing a smooth non-negative spectrum, the shape and width of the peaks are much better reproduced.

\begin{figure}[t]
		\center
		\includegraphics[width=\columnwidth]{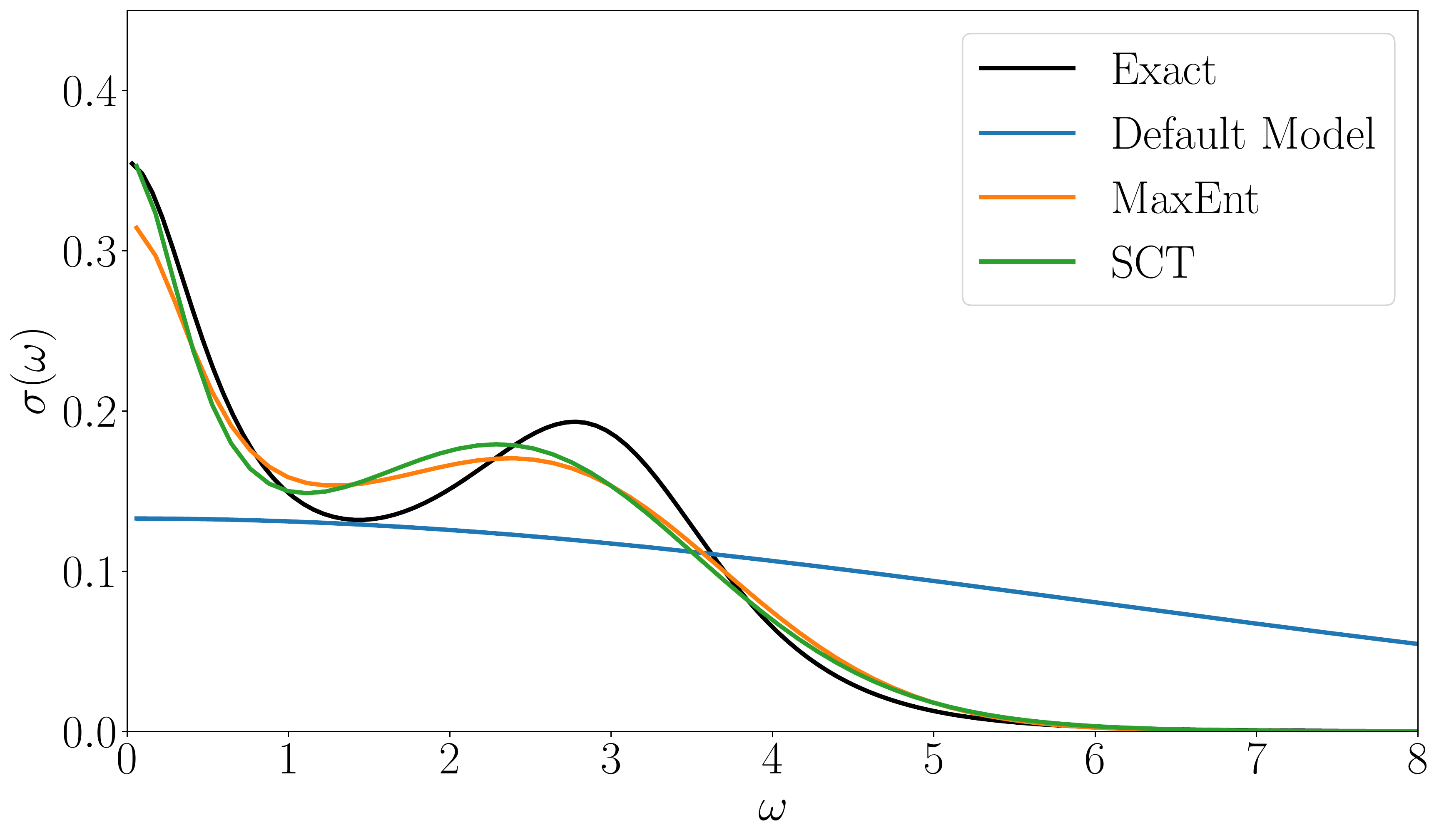}
		\caption{\label{fig:spectra2}
			Comparison of MaxEnt and SCT for a variant of test case 1. 
			This case differs by the location of the second peak and the width of the envelope.
			In the notation of reference~\cite{Ghanem20b}, the optical conductivity used here differs in the values of the following parameters: $\Gamma_e=4, \epsilon_1 = 3$.
			We denote this data set as \emph{test case 2}.
			The default model used here is a scaled Gaussian of width $6$.
		}
	\end{figure}

	In the same plot, we also show the solution of the MaxEnt method using the same starting default model, $\vec{d}^{(0)}$, and a regularization parameter that is also determined by the discrepancy principle.
	Remarkably, the MaxEnt solution is indistinguishably close to SCT solution.
	By examining different other test cases, we have always found that the solutions of MaxEnt and SCT are quite similar and in many cases virtually identical (see Fig.~\ref{fig:spectra2} for another example).
	The following sections will examine and clarify this surprising connection between MaxEnt and SCT. 
	In this context, it is worth noting that MaxEnt has also been recently connected to a specific variant of the average spectrum method, a stochastic method for analytic continuation~\cite{Sandvik22}.
	
	\section{Maximum Entropy Method}
	Similarly to Tikhonov regularization, the Maximum Entropy Method (MaxEnt) introduces a term that penalizes  the mismatch between a spectrum and a default model \cite{Silver90, Jarrell96, Gunnarsson10, Tremblay16}.
	The penalty term, known as  Shannon entropy, is defined as
	\begin{equation}\label{eq:shannon}
		S(\vec{f}|\vec{d}) \coloneqq  \sum_{i=1}^N \left[f_i - d_i - f_i \ln\left(\frac{f_i}{d_i}\right) \right]\;.
	\end{equation}
	It represents the expected amount of information in a spectrum $\vec{f}$ relative to the default model $\vec{d}$.
	This entropy is then optimized in MaxEnt simultaneously alongside the data fit
	\begin{equation}\label{eq:maxent}
		\vec{f}_\text{MaxEnt} (\alpha, \vec{d})\ = \argmax -\frac{1}{2}\chi^2\left(\vec{f}\right) + \alpha S\left(\vec{f}\right|\vec{d})\;.
	\end{equation}
	The fit and entropy trade-off is controlled via the regularization parameter $\alpha$.
	When $\alpha$ is infinitesimally small, MaxEnt formally gives the non-negative least-squares solution, but as $\alpha$ increases, the solution gets smoother and closer to the default model.
	
	There are different ``flavors" of MaxEnt depending on how $\alpha$ is chosen \cite{Jarrell12}.
	The most relevant for our purpose is the one known as Historic MaxEnt.
	In this method, $\alpha$ is chosen such that the fit $\chi^2$ equals the number of the data points $m$.
	This choice is equivalent to the discrepancy principle when the data and the kernel are transformed so that the noise on the data becomes uncorrelated and has unit variance.
	Other commonly-used methods for choosing $\alpha$ are the classical  MaxEnt and Brayn's MaxEnt. 
	Both methods derive a probability distribution over $\alpha$ using Bayesian theory and use either the maximum of this distribution (Classical MaxEnt) or its average (Bryan's MaxEnt) as the final solution.
	In the rest of the paper, we will always assume that the discrepancy principle is applied, and thus, MaxEnt refers to the original way of choosing $\alpha$, i.e.,
	\begin{equation}\label{eq:maxent_disc}
		\vec{f}_\text{MaxEnt}(\vec{d}) =\underset{\vec{f} \in \mathcal{C}} {\argmax}\; S\left(\vec{f}|\vec{d}\right)\;,
	\end{equation}
	where $\mathcal{C}$ is the manifold defined by the discrepancy principle in Eq.~\eqref{eq:disc_set}.
	
	The Shannon entropy is directly related to the Tikhonov regularization term, $T(\vec{f}|\vec{d})$ being the entropy expanded to second order in $\Delta_i \coloneqq f_i - d_i$
	\begin{align}
		S(\vec{f}|\vec{d})
		&=\sum_i \Delta_i-\big(\Delta_i+d_i\big)\ln\left(1{+}\frac{\Delta_i}{d_i}\right)   \nonumber\\
		&\approx  \sum_i \Delta_i  -\frac{\Delta^2_i}{d_i}- d_i \left( \frac{\Delta_i}{d_i} - \frac{\Delta^2_i}{2 d^2_i}\right) \nonumber\\
		&= T(\vec{f}|\vec{d}) \, .
	\end{align}
	This means that the Tikhonov method can be considered an approximation to MaxEnt.
	The quality of this approximation depends on how close the starting default model $\vec{d}$ is to the hypersurface defined by the discrepancy principle $\mathcal{C}$.
	When the default model satisfies the discrepancy principle, then MaxEnt and Tikhonov give the same solution -- the default model itself.
	As the fit of the default model deteriorates, it gets further away from that hypersurface, and the maxima of the penalty terms $S$ and $T$ in $\mathcal{C}$ start to diverge.
	A more quantitative analysis of the difference between MaxEnt and Tikhonov solutions is given in Appendix~\ref{sec:maxent_tikh}.
	
	\section{MaxEnt Family of Equivalent Default Models}
\begin{figure}[t]
		\center
		\includegraphics[width=\columnwidth]{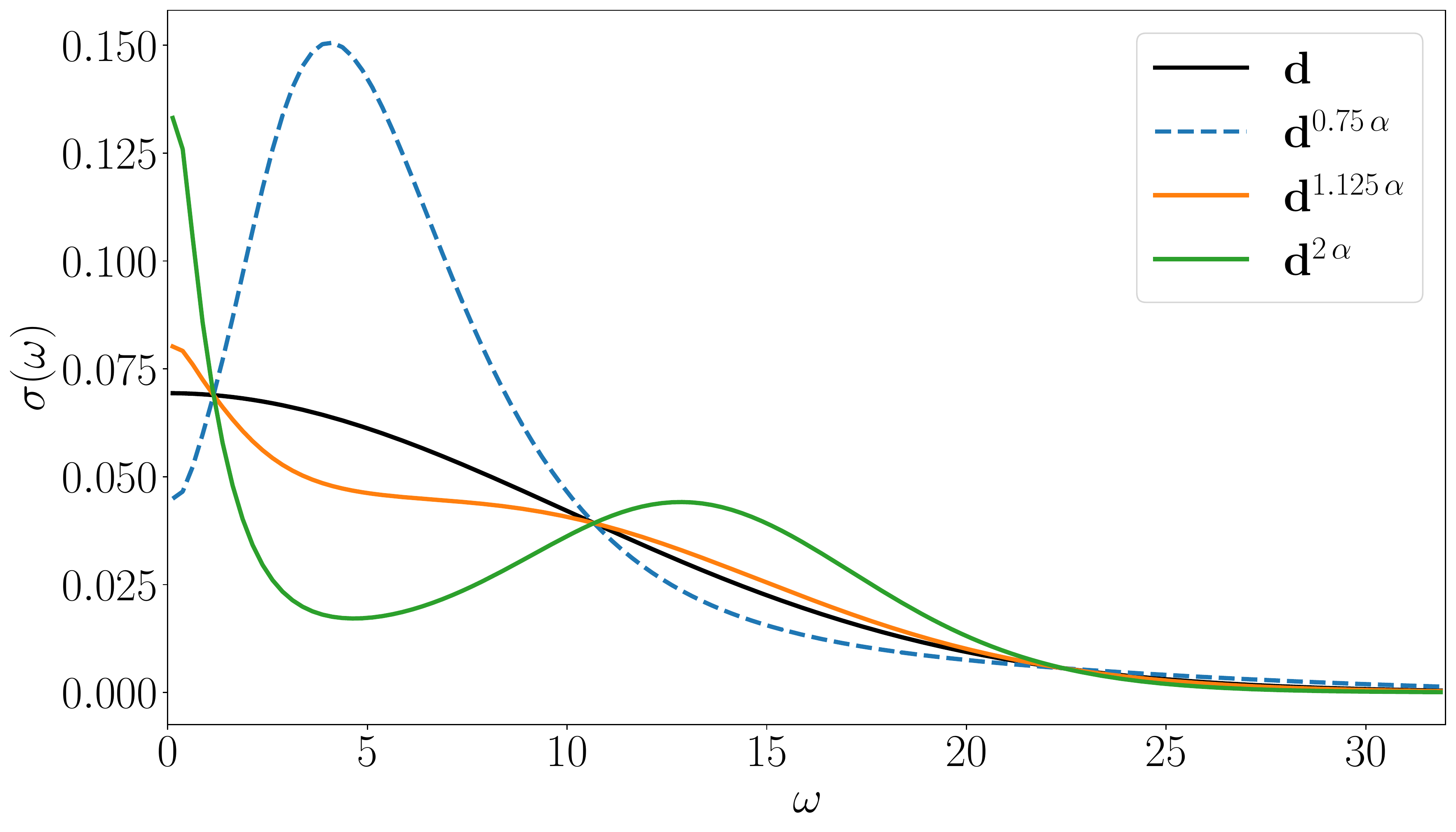}
		\caption{\label{fig:maxent_tikh_dms} 
			Different  default models  equivalent to $\vec{d}$  for test case 1.
			The value of $\alpha$ is determined via the discrepancy principle.}
	\end{figure}

	Analogously to the discussion in Sec.~\ref{sec:discrepancy} about optimizing the Tikhonov penalty, maximizing the Shannon entropy under the discrepancy constraint can also be achieved by following its gradient, projected on $\mathcal{C}$: 
	\begin{equation}
		\vec{b}^{\perp} = \left[ \vec{I}- \frac{ \vec{z}\; \vec{z}^\transpc }{\vec{z}^\transpc\vec{z}}\right] \vec{b} \;,
	\end{equation}
	where $\vec{b} \coloneqq \nabla S$ is the gradient of the entropy with
	\begin{equation}\label{eq:maxent_grad}
		b_i   = - \ln\left( \frac{f_i}{d_i} \right)\;.
	\end{equation}
	At the MaxEnt solution $\vec{f}^\star$, the gradient of Shannon entropy and the gradient of the fit function must be anti-parallel
	\begin{equation}\label{eq:maxent_stationarity}
		\alpha \vec{b}^\star = -\vec{z}^\star\;.
	\end{equation}
	This gives rise to the following self-consistent system of equations satisfied by any MaxEnt solution
	\begin{equation}\label{eq:maxent_consistency}
		f^\star_i = d_i \exp \left( \frac{z_i^\star}{\alpha} \right)\;,
	\end{equation}
	where the fit gradient of the MaxEnt solution $\vec{z}^\star$ depends on the solution itself.
	By rearranging this equation, it becomes clear that the same MaxEnt solution can be obtained using a whole family of other equivalent default models $\vec{d}$ and their corresponding regularization parameters $\alpha$. This family can be constructed explicitly using the MaxEnt solution and its fit gradient:
	\begin{equation}\label{eq:maxent_dm}
		d_i  \coloneqq f^\star_i \exp \left(-\frac{z_i^\star}{\alpha} \right)\;.
	\end{equation}
	Alternatively, given a  default model $\vec{d}$ with regularization parameter $\alpha$, we can construct an entire family of default models $\vec{d}^{\alpha'}$ that result in the same MaxEnt solution $\vec{f}^\star$:
	\begin{equation}\label{eq:maxent_eq_dm}
		d^{\alpha'}_i = d_i \exp \left[- z_i^\star \left(\frac{1}{\alpha'}-\frac{1}{\alpha}\right)\right] .
	\end{equation}
	Note that $\lim_{\alpha'\to\infty} \vec{d}^{\alpha'}=\vec{f}^\star$.
	In Fig.~\ref{fig:maxent_tikh_dms}, we show a set of equivalent default models for test case 1.
	
	Besides establishing the existence of equivalent default models, Eq.~\eqref{eq:maxent_dm} can be used to study the stability of MaxEnt solution with respect to perturbations to these default models.
	The partial derivatives of the default model with respect to variations in MaxEnt solution $\vec{f}^\star$ and regularization parameter $\alpha$ read
	\begin{align}
		\frac{\partial d_i}{\partial f_j^\star} &= \frac{d_i}{f_i^\star} \delta_{i,j} + \frac{d_i}{\alpha}\ \vec{k}_i^\transpc \vec{k}_j\;, \\
		\frac{\partial d_i}{\partial \alpha} &=   \frac{z_i^\star}{\alpha^2} d_i\;.
	\end{align}
	Therefore,  an infinitesimal change in the MaxEnt solution $d\vec{f}^\star$ and   an  infinitesimal change in the regularization parameter $d \alpha$ induce the following relative change in the default model
	\begin{equation}\label{eq:maxent_change}
		\vec{\delta} \coloneqq  \vec{D}^{-1}\  d \vec{d} =  \vec{L} \ d\vec{f}^\star + \frac{d\alpha }{\alpha^2}\ \vec{z}^\star\;,
	\end{equation}
	where $\vec{D}=\text{diag}(\vec{d})$ and $\vec{L}$ is the scaled Hessian of the MaxEnt objective function 
	\begin{equation}\label{eq:maxent_hessian}
		-\alpha\vec{L} \coloneqq -\left(\vec{K}^\transpc\vec{K} +\alpha\vec{F}^\star\right) \;,
	\end{equation}
	with $\vec{F}^\star \coloneqq \text{diag}(\vec{f}^\star)$.
	Inverting Eq.~\eqref{eq:maxent_change} gives the changes in the MaxEnt solution in terms of perturbations to its default model.
	Under the discrepancy principle, the change in the regularization parameter is fixed by the constraint ${\vec{z}^\star}^\transpc d\vec{f}^\star = 0$ (ensuring that $d\vec{f}^\star$ has no component perpendicular to $\mathcal{C}$) to the value
	\begin{equation}
		d\alpha = \alpha^2\ \frac{{\vec{z}^\star}^\transpc \vec{L}^{-1} \vec{\delta}}{{\vec{z}^\star}^\transpc \vec{L}^{-1} \vec{z}^\star}\;,
	\end{equation}
	and the corresponding change in the MaxEnt solution is
	\begin{align}\label{eq:peturb}
		d\vec{f}^\star &=   \vec{L}^{-1}\ \left[\vec{I} -  \frac{\vec{z}^\star {\vec{z}^\star}^\transpc\ \vec{L}^{-1} }{{\vec{z}^\star}^\transpc \vec{L}^{-1} \vec{z}^\star}\ \right]   \vec{\delta}
		=:\vec{L}^{-1}\ \vec{\delta}^{\perp}\;,
	\end{align}
	where $\vec{\delta}^\perp$ is the part of the vector $\vec{\delta}$ perpendicular to the surface normal $\vec{z}^\star$, under the inner product defined by the matrix $\vec{L}^{-1}$.
	
	Relative changes in the default model along the direction of $\vec{z}^\star$ give an equivalent default model and thus have no effect on the MaxEnt solution. To assess the effect of changes in the default model along orthogonal directions, we need to look into the spectral decomposition of the matrix $\vec{L}$.
	The eigenvectors of  $\vec{L}$ match the spectral modes of the rescaled kernel $\vec{K}^\prime \coloneqq \vec{K} \sqrt{\vec{F}^\star}$ and the eigenvalues of the former $\lambda_i$ are related to the singular values of the later $s^\prime_i$ as
	\begin{equation}
		\lambda_i = \frac{\alpha+{s^\prime}_i^2}{\alpha}\;.
	\end{equation}
	We now distinguish two limiting cases depending on the direction of the vector $\vec{\delta}^\perp$.
	When ${s}'^2_i \gg \alpha$, then $\lambda^{-1}_i \approx \alpha/{s}_i'^{2}$. Therefore, changes along the leading modes have little effect on the MaxEnt solution, and the effect is smaller the further away  the default model is  from $\mathcal{C}$.
	On the other hand, when ${s}'^2_i \ll \alpha$, then $\lambda^{-1}_i \approx 1$. Therefore, changes along the trailing modes are directly reflected in the MaxEnt solution.
	Assuming that a MaxEnt solution is smooth, the leading modes of $\vec{K}^\prime$ are smooth and slowly varying functions while the trailing ones are highly oscillating. These results then confirm and elucidate the common wisdom that slowly varying details of the default model have little to no effect on MaxEnt solutions, while sharp features tend to introduce strong biases.
	Finally, note that having more accurate data scales up the singular values $s_i^\prime$, and thus MaxEnt solution becomes less sensitive to changes in the default model, as one would intuitively anticipate.
	
	\section{Connecting  SCT to MaxEnt}
	Let $\vec{d}^{(t)}$ be the default model at step $t$ of SCT and $\vec{f}^{(t)}$ and $\vec{z}^{(t)}$ be the corresponding Tikhonov solution and its fit gradient.
	By combining Eq.~\eqref{eq:tikh_grad} with Eq.~\eqref{eq:tikh_stationarity},  we see that the Tikhonov solutions satisfy the following self-consistent equation (analogous to Eq.~\eqref{eq:maxent_consistency} of MaxEnt)
	\begin{equation}\label{eq:tikh_consistency}
		f^{(t)}_i = d^{(t)}_i  \left[1 +\frac{ \mathbf{z}^{(t)}}{\alpha^{(t)}} \right]\;.
	\end{equation}
	Using mixing parameters $\mu^{(t)}$, the default models at subsequent iterations are then related by
	\begin{multline}\label{eq:sct_mixing}
		d_i^{(t+1)} = \big(1{-}\mu^{(t)}\big) d_i^{(t) }+ \mu^{(t)} f_i^{(t)}
		=  d_i^{(t)} \left[1 +\frac{\mu^{(t)}}{\alpha^{(t)}} \ \vec{z}^{(t)} \right].
	\end{multline}
	Applying this relation recursively and assuming very small $\mu^{(t)}/\alpha^{(t)}$, we get the following exponential form for the default models produced by SCT
	\begin{equation}\label{eq:sct_exp}
		d_i^{(t)}  = d^{(0)}_i \exp\left[ \sum_{\tau=0}^{t-1} \frac{\mu^{(\tau)} }{\alpha^{(\tau)}}\  \vec{z}^{(\tau)}\right]
		= d^{(0)}_i  \exp \left[\frac{ \vec{\tilde{z}}^{(t)}}{\tilde{\alpha}^{(t)}} \right]\;,
	\end{equation}
	where in the last equation we defined the effective fit gradients $\vec{\tilde{z}}^{(t)}$ and the effective regularization parameters $\tilde{\alpha}^{(t)}$ as 
	\begin{equation}
		\vec{\tilde{z}}^{(t)} \coloneqq  \tilde{\alpha}^{(t)}  \sum_{\tau=0}^{t-1} \frac{\mu^{(\tau)}}{\alpha^{(\tau)}}  \mathbf{z}^{(\tau)} , \quad \frac{1}{\tilde{\alpha}^{(t)}} \coloneqq  \sum_{\tau=0}^{t-1} \frac{\mu^{(\tau)}}{\alpha^{(\tau)}}\;.
	\end{equation}
	Comparing the default models generated by SCT [cf. Eq.~\eqref{eq:sct_exp}] with the MaxEnt family of equivalent default models  [cf. Eq.~\eqref{eq:maxent_eq_dm}], it is clear that the two have the same functional form and would match if the effective fit gradients $\vec{\tilde{z}}^{(t)}$ match the MaxEnt fit gradient $\vec{z}^\star$. 
	
\begin{figure}[t]
		\center
		\includegraphics[width=\columnwidth]{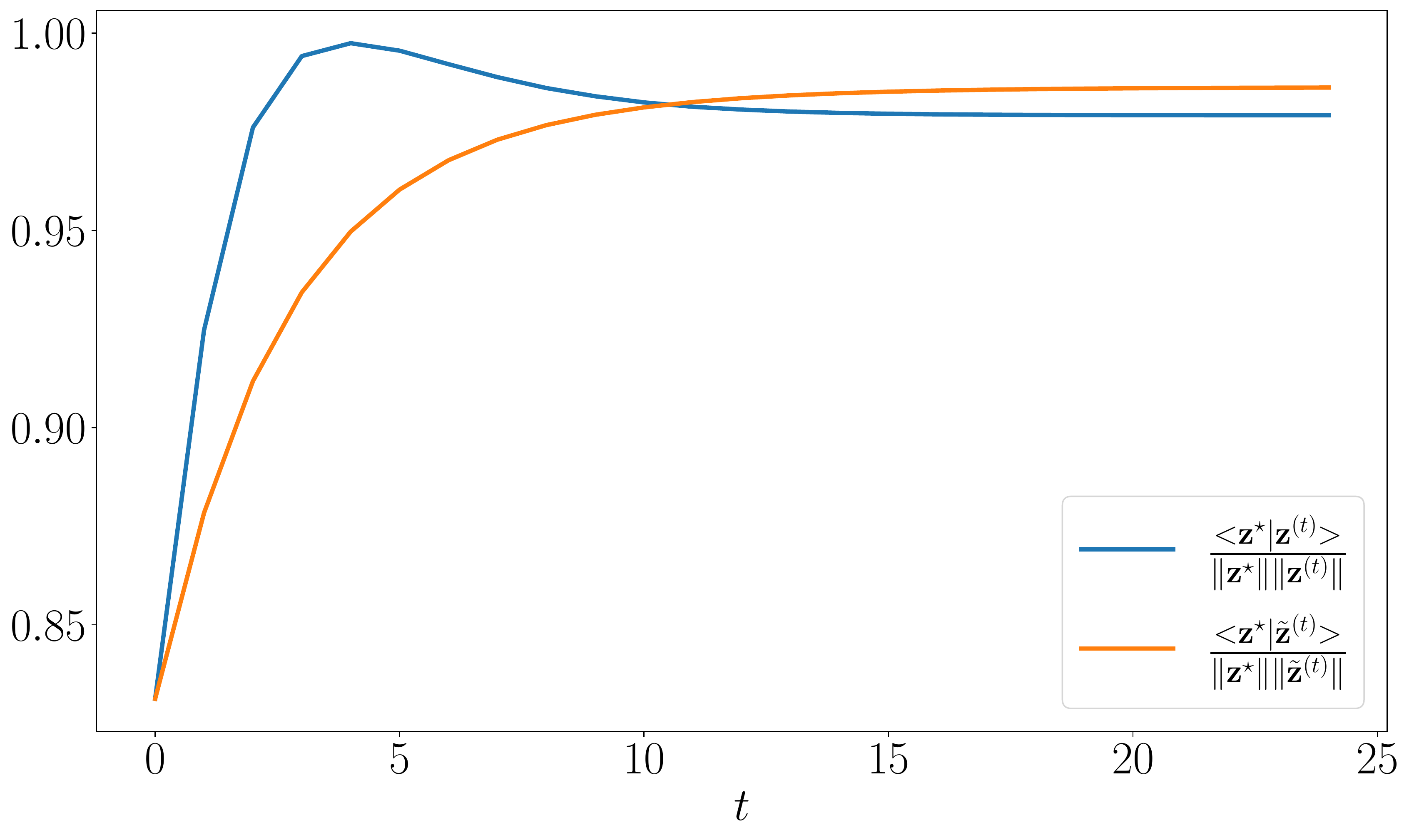}
		\caption{\label{fig:maxent_sct_z_overlap} 
			Normalized overlap between the MaxEnt fit gradient $\vec{z}^\star$ and the fit gradients produced at different SCT iterations (denoted as $t$) in test case 1.
			Both the bare gradients $\vec{z}^{(t)}$ (gradients of Tikhonov solutions) and the effective gradients $\tilde{\vec{z}}^{(t)}$ are shown.
			The overlaps and norms are calculated using the inner product $\braket{\vec{x}, \vec{y}} \coloneqq \vec{x}^\transpc \vec{L}^{-1} \vec{y}$, where $\vec{L}$ is the scaled Hessian of MaxEnt objective function defined in Eq.~\eqref{eq:maxent_hessian}.
		}
	\end{figure}

\begin{figure}[t]
		\center
		\includegraphics[width=\columnwidth]{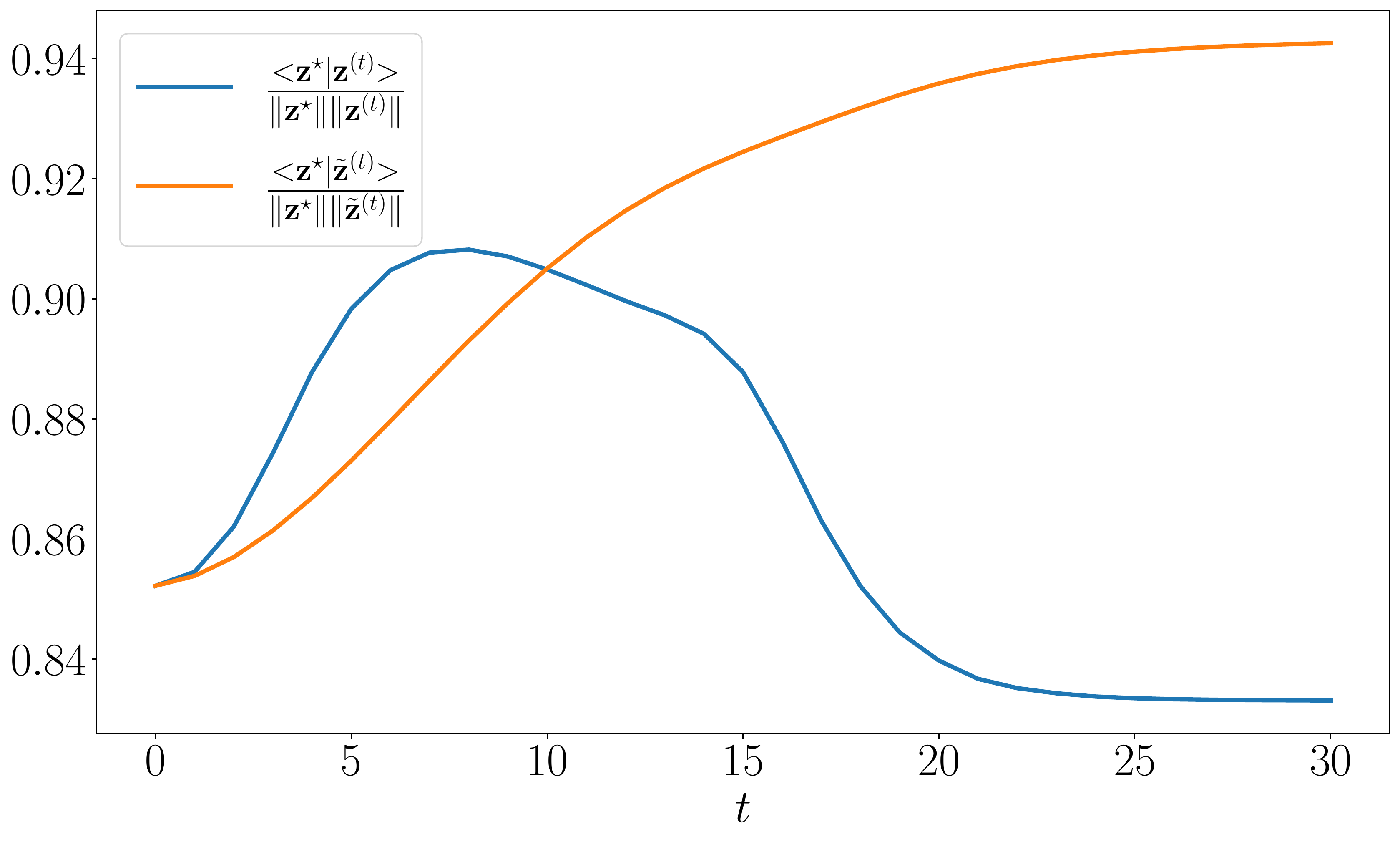}
		\caption{\label{fig:overlap2} 
			Normalized overlap between the MaxEnt fit gradient $\vec{z}^\star$ and the fit gradients produced at different SCT iterations (denoted as $t$) in test case 2.
		}
	\end{figure}

	Indeed, the effective gradients of SCT provide an excellent approximation to the MaxEnt gradient. In Fig.~\ref{fig:maxent_sct_z_overlap}, we plot the normalized overlap between the two at different iterations of SCT. 
	The starting effective gradient is nothing but the original Tikhonov gradient, which already has a very good overlap of $0.83$.
	This is to be expected since,  as discussed in the previous section, Tikhonov provides an approximation to MaxEnt.
	As the SCT procedure iterates, the effective gradient not only maintains the good initial overlap, but the overlap improves until it saturates at about $0.99$ when the procedure converges.
	Interestingly, the overlap with the ``bare" gradients $\vec{z}^{(t)}$, i.e., the gradients of Tikhonov solutions at different iterations, does not necessarily increase.
	The plot shows that the bare overlap actually drops after a couple of iterations.
	We observed cases where the bare overlap even drops below its starting value (see Fig.~\ref{fig:overlap2}).
	Nevertheless, in all cases we investigated, the effective gradients always had a monotonically-increasing overlap with the MaxEnt gradient. An argument for this behavior of the fit gradients is detailed in Appendix~\ref{sec:sct_stability}.
	
	These results demonstrate that the set of default models produced by SCT provides an  approximation to the MaxEnt family of equivalent default models, and thus solving the MaxEnt problem with any one of them gives a solution that is close to the solution of the original MaxEnt problem.
	At convergence, the default model of SCT satisfies the discrepancy principle, and thus, it is trivially the solution of its own MaxEnt problem and a good approximation of the original MaxEnt solution.
	In Appendix~\ref{sec:newton}, we give an alternative perspective in which SCT can be seen as an approximate and simplified variant of Newton's method for obtaining the MaxEnt solution.
	
	\section{Summary}
	In this paper, we used singular value decomposition to derive a generally-applicable method for estimating the noise level on QMC data. Having a reliable error estimate is crucial when using the discrepancy principle/Historic MaxEnt.
	We then introduced a particular form of the Tikhonov regularization that is more suitable for analytic continuation problems. 
	Besides solving the implicit grid dependence and normalization issues, this form is closely connected to Shannon entropy.
	A quadratic approximation of the entropy around its default model gives precisely the introduced Tikhonov penalty term.
	This form allows approximating the MaxEnt solution using the Tikhonov method when the default model already has a good fit to the data (i.e., in the limit of large regularization parameter).
	In the typical cases where the default model does not fit the data well, we showed that an iterative procedure where the default model is repeatedly mixed with its Tikhonov solution still gives similar results to MaxEnt.
	We investigated the connection between the two methods, which revealed that the same MaxEnt solution could be produced by a whole family of equivalent default models. 
	This family is approximately traced by the the self-consistent Tikhonov procedure.
	SCT  provides a simple and efficient alternative to MaxEnt that could be easily implemented using any linear algebra library.
	In particular, we expect SCT to be useful for the analytic continuation of matrix-valued Green functions, where MaxEnt is trickier to implement~\cite{Kraberger17, Jiani21}.
	
	\appendix
	
	\section{Tikhonov solution using SVD}\label{app:tikh}
	The minimization problem of Tikhonov in Eq.~\eqref{eq:tikh} can be written as the following least squares problem with an extended kernel matrix and extended data vector
	\begin{equation}
		\vec{f}_\text{Tikhonov}(\alpha, \vec{d}) = \underset{\vec{f}}{\argmin} \norm{
			\begin{pmatrix} \vec {K} \\ \sqrt{\alpha\ \vec{D}^{-1}}
			\end{pmatrix}  \vec{f} - \begin{pmatrix} \vec {g} \\ \sqrt{\alpha \vec{D} }\vec{e}
		\end{pmatrix}  }^2\;.
	\end{equation}  
	where $\vec{D}=\text{diag}(\vec{d})$ and $\vec{e}=\left(1, 1, \dots, 1\right)^\transp$.
	The normal equation of this least-squares problem reads
	\begin{align}\label{eq:tikh_normal}
		\left(\vec{K}^\transp\mathbf{K} + \alpha \vec{D}^{-1}\right) \vec{f} &= \vec{K}^\transp\vec{g}+\alpha \vec{e}  \\ \Leftrightarrow
		\left[\vec{\tilde{K}}^\transp\vec{\tilde{K}} + \alpha \vec{I}\right]\left(\sqrt{\vec{D}^{-1}}\ \mathbf{f}\right) &= \vec{\tilde{K}}^\transp\mathbf{g}+\alpha\sqrt{\vec{D}}\vec{e}\;,
	\end{align}
	where a rescaled kernel matrix $\vec{\tilde{K}} $  is defined as $ \vec{\tilde{K}} \coloneqq \vec{K} \sqrt{\vec{D}}  $.
	Using SVD of the rescaled matrix $ \vec{\tilde{K}} = \vec{\tilde{U}} \vec{\tilde{S}} \vec{\tilde{V}}^\transp$, the normal equation in the mode space reads
	\begin{equation}
		\left[ \vec{\tilde{S}}^\transp \vec{\tilde{S}} + \alpha \vec{I} \right] \vec{\tilde{V}} ^\transp  \left(\sqrt{\vec{D}^{-1}}\ \mathbf{f}\right)  = \vec{\tilde{S}} \vec{\tilde{U}}^\transpc \vec{g} + \alpha \vec{\tilde{V}}^\transp \sqrt{\vec{D}} \vec{e}\;.
	\end{equation}
	The Tikhonov solution can then be expressed in terms of the rescaled modes of the rescaled matrix $\vec{V}^\prime \coloneqq \sqrt{\vec{D}} \vec{\tilde{V}}$  as
	\begin{equation}\label{eq:tikh_expansion}
		\vec{f}_\text{Tikhonov} = \sum_i \frac{\tilde{s_i}^2}{\tilde{s_i}^2+\alpha} \ \frac{\vec{\tilde{u}}_i^\transpc \vec{g}}{\tilde{s_i}} \vec{v}^\prime_i + \alpha \sum_i \frac{{\vec{v}^\prime_i}^\transp \vec{e}}{\tilde{s_i}^2+\alpha}  \vec{v}^\prime_i \;.
	\end{equation}
	The first term is similar to the expansion of the original grid-dependent Tikhonov in Eq.~\eqref{eq:btikh_svd}, while the second term comes from centering the regularization term around the default model.
	
	Unlike the spectral modes $\vec{v_i}$ in Eq.~\eqref{eq:btikh_svd}, however, the modes $\vec{v^\prime_i}$ are not orthonormal under the standard inner product.
	They are instead orthonormal under the modified inner product
	\begin{equation}
		\braket{\vec{x}, \vec{y}} \coloneqq \vec{x}^\transp \vec{D}^{-1} \vec{y}\;.
	\end{equation}
	Moreover, these vectors can be seen as the modes of the original kernel matrix, with the orthogonality being defined under this modified inner product. This view holds since
	\begin{equation}
		\vec{K} \vec{v}^\prime_i = \vec{\tilde{K}} \vec{\tilde{v}}_i = \tilde{s}_i \vec{\tilde{u}}_i \;, 
	\end{equation}
	and
	\begin{equation}
		\braket{\vec{v}^\prime_i, \vec{v}^\prime_j} =  \delta_{i,j}\;.
	\end{equation}
	Eq.~\eqref{eq:tikh_expansion} can then be seen as a direct expansion of the Tikhonov solution in terms of the spectral modes of the kernel matrix
	\begin{equation}
		\vec{f}_\text{Tikhonov} = \sum_i \braket{\vec{v}^\prime_i, \vec{f}_\text{Tikhonov}}  \vec{v}^\prime_i\;,
	\end{equation}
	with 
	\begin{equation}
		\braket{\vec{v}^\prime_i, \vec{f}_\text{Tikhonov}} =  \frac{1}{\tilde{s_i}^2+\alpha} \left[\tilde{s}^2_i\ \frac{\vec{\tilde{u}}_i^\transpc \vec{g}}{\tilde{s}_i} + \alpha \braket{\vec{v}^\prime_i, \vec{d}} \right]\;.
	\end{equation}
	Note how each component of the Tikhonov solution is an interpolation between the components of the least squares spectrum and the default model.
	Different components, however, are mixed differently (each according to its singular value), and thus the overall Tikhonov solution is generally not a simple interpolation of the two spectra.
	
	\section{Difference between MaxEnt and Tikhonov}\label{sec:maxent_tikh}
	We can quantify the difference between the Tikhonov and MaxEnt solutions of the same default model and regularization parameter as following
	\begin{equation}
		\vec{\Delta}^\star \coloneqq \vec{f}_\text{Tikhonov}- \vec{f}^\star =  \vec{H}^{-1} \ \vec{\nabla}T^{\vec{f}^\star}\;,
	\end{equation}
	where $\vec{H}$ is minus the Hessian  of the Tikhonov objective function of Eq.~\eqref{eq:tikh}
	\begin{equation}\label{eq:hessian}
		\vec{H} \coloneqq {\alpha}\ \vec{D}^{-1} + \vec{K}^\transpc\vec{K}\;,
	\end{equation}
	and $\vec{\nabla}T^{\vec{f}^\star}$ is its gradient at the MaxEnt solution
	\begin{align}
		\vec{\nabla}T^{\vec{f}^\star}_i &=z_i^\star + {\alpha}\ a_i^\star = z_i^\star - {\alpha}\ \frac{f^\star_i-d_i}{d_i} \nonumber \\
		&= z^\star_i - {\alpha} \left[\exp \left({\frac{z_i^\star}{\alpha}} \right)- 1 \right] \nonumber \\
		& =  - \frac{1}{2 \alpha} {z_i^\star}^2 + \mathcal{O}(\alpha^{-2} )\;.
	\end{align}
	Therefore, the gradient scales linearly with the inverse of $\alpha$. 
	To analyze how the difference $\vec{\Delta}^\star$ scales, we look at the spectral decomposition of the Hessian matrix $\vec{H}$.
	Its eigenvectors are the same as the spectral modes of the rescaled matrix $ \vec{\tilde{K}} = \vec{K} \sqrt{\vec{D}}$, and its eigenvalues $h_i$ are related to the singular values of $\vec{\tilde{K}} $ as following
	\begin{equation}
		h_i = {\alpha} + {\tilde{s}_i}^2\;.
	\end{equation}
	The $i$-th component of the difference then scales as $1/(\alpha^2+\alpha \tilde{s}_i^2)$, and thus, the difference between Tikhonov and MaxEnt vanishes quadratically in the limit of strong regularization.
	Note that the components of the gradient along the leading spectral modes (i.e., the smooth components with large singular values ) get suppressed more than the trailing ones (i.e., the oscillating components with small singular values).
	
	\section{Dynamics of SCT}\label{sec:sct_stability}
	Let $\vec{z}^\prime$ be the fit gradient of a Tikhonov solution.
	When mixing the default model with the Tikhonov solution, the relative change in the default model is proportional to this fit gradient, namely $\vec{\delta} = \mu / \alpha\ \vec{z}^\prime$.
	The part of $\vec{z}^\prime$ along the MaxEnt fit gradient   $\vec{z}^\star$ gives an equivalent default model and thus does not affect the MaxEnt solution.
	Let $d \vec{z}^\prime$ denote the part of $\vec{z}^\prime$ perpendicular to $\vec{z}^\star$ under the inner product defined by $\vec{L}^{-1}$ i.e.
	\begin{equation}
		d \vec{z}^\prime \coloneqq \vec{z}^\prime - \frac{{\vec{z}^\star}^\transpc \vec{L}^{-1} \vec{z}^\prime}{{\vec{z}^\star}^\transpc \vec{L}^{-1} \vec{z}^\star}\ \vec{z}^\star\;.
	\end{equation}
	Then the relevant relative change in the default model is $\vec{\delta}^\perp  =\mu/\alpha\ d \vec{z}^\prime$.
	From Eq.~\eqref{eq:peturb}, we see that the corresponding change in the fit gradient of the MaxEnt solution reads
	\begin{align}\label{eq:stability}
		d\vec{z}^\star &=  -\vec{K}^\transpc\vec{K}\ d\vec{f}^\star =    -\vec{K}^\transpc\vec{K}\ \vec{L}^{-1}\ \vec{\delta}^{\perp} \nonumber \\
		&= -\frac{\mu}{\alpha} \ \vec{K}^\transpc\vec{K}\ \vec{L}^{-1}\ d\vec{z}^\prime\;.
	\end{align}
	Given that the matrices  $\vec{L}^{-1}$ and  $\vec{K}^\transpc\vec{K}$ are positive semi-definite, the overlap between $d\vec{z}^\star $ and $d\vec{z}^\prime$ is non-positive, i.e., the fit gradient of the MaxEnt solution moves opposite to the change in the fit gradient that induced it. 
	Since Tikhonov solutions generally follow the MaxEnt solutions, the new Tikhonov gradients would be closer to the original MaxEnt gradient than the previous ones.
	This explains why the bare fit gradient vectors in SCT initially move closer to the original MaxEnt gradient vector (Figs.~\ref{fig:maxent_sct_z_overlap} and \ref{fig:overlap2}).
	However, the MaxEnt solutions using SCT default models keep drifting away in the same direction, so the Tikhonov solutions and their fit gradients would eventually also start moving away from the original MaxEnt.
	The effective fit gradient, on the other hand, is an average of these bare gradients and thus can be closer to the original MaxEnt than any of its summands. This happens when the bare gradients circulate around the original MaxEnt gradient, which is the case in SCT.
	
\begin{figure}[t]
		\center
		\includegraphics[width=\columnwidth]{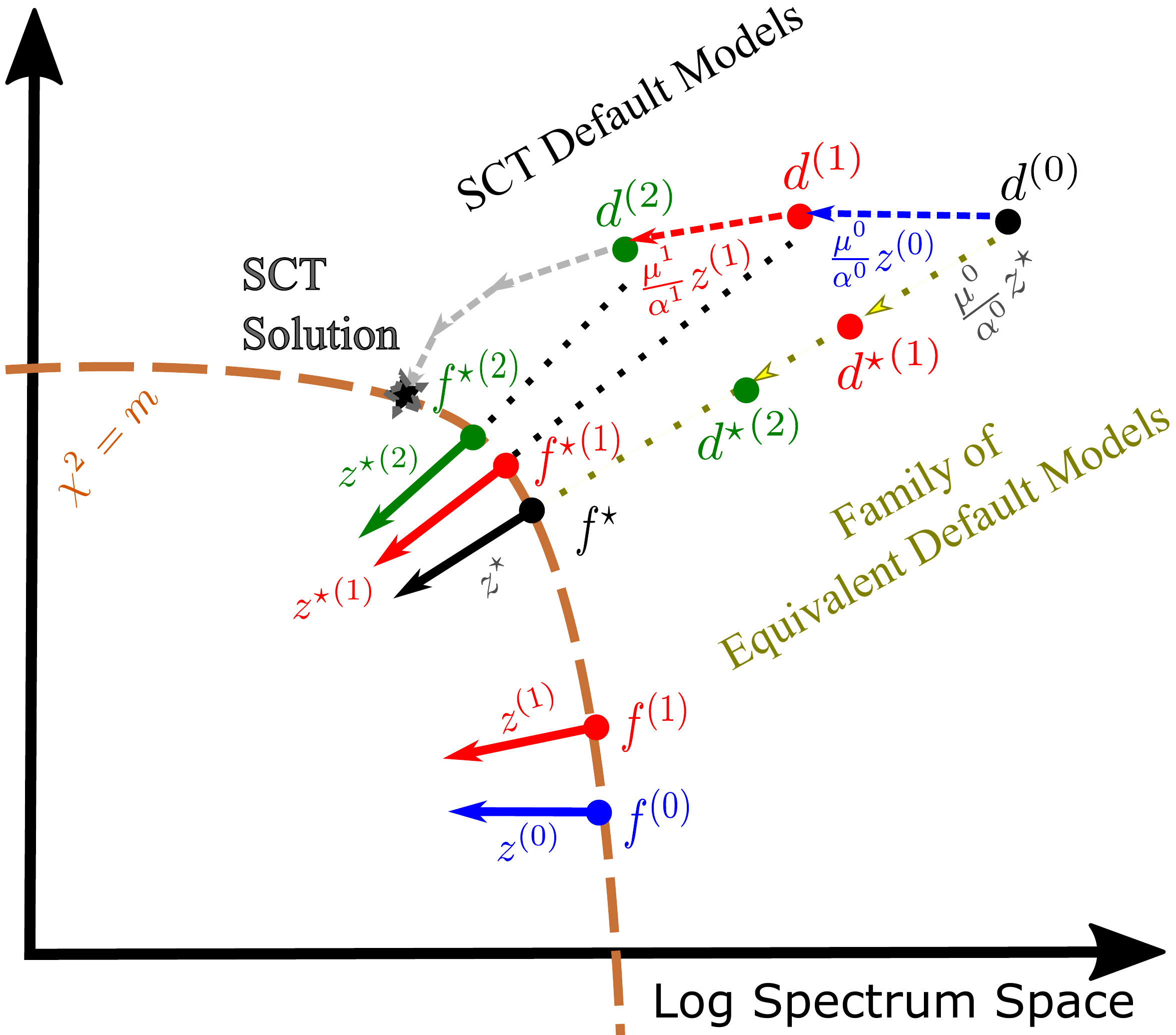}
		\caption{\label{fig:drawing} 
			Schematic diagram illustrating how the default models and their MaxEnt and Tikhonov solutions evolve with the SCT iterations.
			The diagram is depicted in the logarithmic space of spectra.
			Note that Tikhonov solutions are assumed here to be strictly positive, although, in general, they may have either sign.
		}
	\end{figure}

	The dynamics described above is depicted schematically in Fig.~\ref{fig:drawing}.
	In this diagram, we represent the log spectra as points, so the family of equivalent default models ${\vec{d}^\star}^{(1)}, {\vec{d}^\star}^{(2)}, \dots$ all lie on a straight line between the initial default model $\vec{d}^{(0)}$ and its MaxEnt solution ${\vec{f}^\star}$. This line is specified by the fit gradient vector $\vec{z}^\star$.
	In SCT,  $\vec{z}^\star$ is replaced by $\vec{z}^{(t)}$, the bare fit gradients at the Tikhonov solutions $\vec{f}^{(t)}$, leading to a set of alternative default models $\vec{d}^{(t)}$ that approximates the equivalent family ${\vec{d}^\star}^{(t)}$.
	Each approximate default model $\vec{d}^{(t)}$ has its own MaxEnt solution ${\vec{f}^\star}^{(t)}$ and Tikhonov solution  ${\vec{f}}^{(t)}$.
	In this two-dimensional case, according to Eq.~\eqref{eq:stability}, the Tikhonov solution ${\vec{f}}^{(t)}$ and the MaxEnt solution at the next iteration  ${\vec{f}^\star}^{(t+1)}$ must be on opposite sides of the MaxEnt solution ${\vec{f}^\star}^{(t)}$.
	Therefore, the fit gradients of Tikhonov $\vec{z}^{(t)}$ would initially get closer to $\vec{z}^\star$ before moving away.
	Also, note how the effective fit gradients (i.e., consecutive weighted averages of $\vec{z}^{(t)}$)   get monotonically closer to  $\vec{z}^\star$.
	This is the result of $\vec{z}^{(t)}$ moving from one side of $\vec{z}^\star$ to the other, and the weights $\mu^{(t)}/\alpha^{(t)}$ getting lower for higher iterations. 
	\section{SCT as Reset Newton Method}\label{sec:newton}
	Another perspective on SCT is seeing it as a variant of Newton's method for optimization. 
	Assuming that the optimal regularization parameter for satisfying the discrepancy principle is somehow known in advance, solving the MaxEnt problem of Eq.~\eqref{eq:maxent_disc} reduces to optimizing the MaxEnt objective function of Eq.~\eqref{eq:maxent}. Using the default model $\vec{d}$ as an initial guess, an improved solution can be obtained using Newton's method as
	\begin{equation}\label{eq:newton}
		\vec{d}^\prime = \vec{d} + \gamma \vec{H}^{-1} \ \vec{\nabla}S^{\vec{d}}\;,
	\end{equation}
	where $\gamma$ is a small step size and $\vec{H}$ is minus the Hessian of the objective function at the default model (which coincides with minus the Tikhonov Hessian in Eq.~\eqref{eq:hessian}) and $\vec{\nabla}S^{\vec{d}}$ is its gradient, also evaluated at the default model.
	
	The vector  $\vec{x} \coloneqq \vec{H}^{-1} \vec{\nabla}S^{\vec{d}}$ is the solution of 
	\begin{equation}
		\begin{split}
			\left[ \vec{K}^\transpc\vec{K}+{\alpha}\ \vec{D}^{-1}\right] \vec{x}  &= \vec{K}^\transpc \left[\vec{g} - \vec{K} \vec{d}\right] \\ 
			\Leftrightarrow  \left[\vec{K}^\transpc\vec{K} + {\alpha}\ \vec{D}^{-1} \right]  \left[\vec{x}+\vec{d}\right] &= \vec{K}^\transpc \vec{g} + \alpha \vec{e} \;.
		\end{split}
	\end{equation}
	Comparing with Eq.~\eqref{eq:tikh_normal}, we see that  $\vec{x} + \vec{d}$ equals the Tikhonov solution, and thus Newton's update formula can be written as
	\begin{equation}
		\vec{d}^\prime = \vec{d} + \gamma \big(\vec{f}_\text{Tikhonov} - \vec{d}\big)\;,
	\end{equation}
	which is precisely the mixing formula used in SCT.
	Note that the entropy has no contribution to the gradient vector $\vec{\nabla}S^{\vec{d}}$ at the starting default model.
	However, at later steps there is an additional term $-\alpha^{(t)} \ln({d^{(t)}_i}/{d^{(0)}_i})$.
	SCT ignores this term; thus, SCT is equivalent to Newton's method, where the default model is always reset to its most recent solution.
	
	Interestingly, the missing entropy contributions can be expressed in terms of the effective fit gradients 
	\begin{equation}
		-\alpha^{(t)} \ln\left(\frac{d^{(t)}_i}{d^{(0)}_i}\right) = - \frac{\alpha^{(t)}}{\tilde{\alpha}^{(t)}} \vec{\tilde{z}}^{(t)}\;.
	\end{equation}
	Therefore, we can recover the full Newton's method as a variant of the SCT method where the data is modified at each step to take into account the residuals of the previous Tikhonov solutions.
	
	\bibliography{manuscript}
\end{document}